\documentclass[11pt]{article}

\usepackage{amsthm}
\usepackage{amssymb}
\usepackage{amsmath}
\newcommand{\myif}{\quad\text{if}\quad}

\DeclareMathAlphabet\EuFrak{U}{euf}{m}{n}       
\SetMathAlphabet\EuFrak{bold}{U}{euf}{b}{n}     

\begin{document}
\author{Sergio Doplicher
                         \\Dipartimento di Matematica
                         \\University of Rome ``La Sapienza''
                         \\00185 Roma, Italy  }

\title{Spin and Statistics and First Principles}
\maketitle

\begin{abstract} It was shown in the early Seventies that, in Local 
Quantum Theory (that is the most general formulation of Quantum Field 
Theory, if we leave out only the unknown scenario of Quantum Gravity) the 
notion of Statistics can be grounded solely on the local observable 
quantities (without assuming neither the commutation relations nor even 
the existence of unobservable charged field operators); one finds that 
only the well known (para)statistics of Bose/Fermi type are allowed by 
the key principle of local commutativity of observables. In this frame it 
was possible to formulate and prove the Spin and Statistics Theorem 
purely on the basis of First Principles.

In a subsequent stage it has been possible to {\itshape prove} the 
existence of a 
unique, canonical algebra of local field operators obeying ordinary 
Bose/Fermi commutation relations at spacelike separations.

In this general guise the Spin - Statistics Theorem applies to Theories 
(on the four dimensional Minkowski space) where only massive particles 
with finite mass degeneracy can occur. Here we describe the underlying 
simple basic ideas, and briefly mention the subsequent generalisations; 
eventually we comment on the possible validity of the Spin - Statistics 
Theorem in presence of massless particles, or of violations of locality 
as expected in Quantum Gravity.

  \end{abstract}

\section{What is Statistics?} Most would answer: ``look at the 
commutation/anticommutation relations between spacelike separated field 
operators''; but field operators usually are {\itshape not} observable 
quantities, 
their properties might be mere features of the formalism
- or they might even fail to exist.

Otherwise, many would answer: ``Construct n particle states, then take the 
symmetric/antisymmetric part of the tensor product, then\ldots''. 

But this means 
{\itshape to impose} a choice in the construction of free field operators; to 
do 
something similar in an generic interacting theory, we must:

(i)  prove that there exist a product operation between suitably 
localised states in the theory which produces other states in the 
{\itshape same} 
theory (i.e. expectation functionals on the algebra of observable 
quantities), which describe the composed states. Such composed states 
should be independent of the order in which the factors are listed.

(ii) associate to that  product of states in a canonical way a product  
state vector: the latter may depend upon the order, e.g. changing under a 
permutation of factors by a phase or by a unitary operator which commutes 
with all observables.

(iii) if all the factors are vector states of a fixed superselection 
sector, the statistics of that superselection sector can then be defined 
by the actions of the permutation groups (of all possible orders for the 
different numbers of factors) obtained as above (provided we can show 
that the way the product state vector changes under permutations of the 
order of factors, is indeed described by an action of the permutation 
group which depends only upon the choice of the superselection sector).

Note that the distinction between integer/half integer spin arises in a 
similar way: rotations of $2\pi$ leave expectation functional (states) 
invariant, but may change the phase of the state vectors (only by a sign 
if the choice is canonical).

It is a remarkable fact that the above approach to statistics can be 
realized in any Local Quantum Theory, based essentially only on the 
Locality Principle.

More precisely, suppose the observables are given as bounded operators on 
a fixed Hilbert space, describing a single superselection sector, 
{\itshape the vacuum sector}. Their collection is therefore 
irreducible. The main postulate is that this is the collection of 
{\itshape(quasi) 
local observables} \cite{dopl1,dopl2}, i.e. we have an inclusion preserving map from nice 
regions (say the set of double cones - the intersections of open forward 
and backward light cones with a common interior point) in Minkowski space 
to * subalgebras of operators 
\begin{equation}
\label{dopleq1}
\mathcal O  \mapsto  \mathfrak A(\mathcal O) \subset B(\mathcal H)
\end{equation}
whose selfadjoint elements are the observables which can be measured in 
the spacetime region $\mathcal O$, and such that {\itshape local 
commutativity} holds, i.e. the 
measurements of two spacelike separated observables must be compatible, 
so that they commute with each other:
\begin{equation}
\label{dopleq2}
\mathfrak A(\mathcal O_1) \subset \mathfrak A(\mathcal O_2)' \myif \mathcal O_1 
\subset \mathcal O_2'                                      
\end{equation}
where the prime on a set of operators denotes its commutant (the set of 
all bounded operators commuting with all the operators in the given set) 
and on a set in Minkowski space denotes the spacelike complement. Thus 
each $\mathfrak A(\mathcal O)$ is included in the intersection of the commutants of all $\mathfrak A(\mathcal O_n)$, 
as $O_n$ runs through all the double cones spacelike to $\mathcal O$.

This axiom is strengthened to  {\itshape Duality}: each \(\mathfrak A(\mathcal 
O)\) is maximal with the above 
property, namely that inclusion is actually an equality:
\begin{equation} \label{dopleq3} \mathfrak A(\mathcal O)  = \mathfrak 
A(\mathcal O')',  \end{equation}
where, here and in the following, $\mathfrak A(\mathcal O')$ denotes the norm 
closed *subalgebra 
generated by all the local algebras associated to the various double cones 
which are spacelike separated from $\mathcal O$, i.e. included in $ \mathcal 
O'$.

(A weaker form of this assumption is ``essential duality'', requiring only 
that the 
$\mathfrak A(\mathcal O')'$ locally commute with one another; if the theory 
is suitably 
described by Wightman fields,  essential duality can be proved to hold 
 \cite{dopl3} ; the weakening of duality to essential duality indicate the 
presence of spontaneously broken global gauge symmetries \cite{dopl4} ).

Translation and Lorentz covariance, Spectrum Condition play no role in
this analysis, except for a mild technical 
consequence, proven long ago by
Borchers , that we called {\itshape the Property B}\footnote{
{\itshape Property B}: If $\mathcal O_1$ and $\mathcal O_2$
are double cones and the second includes the closure of the
first, then any selfadjoint projection $E$  localised in the first is of the
form $E  =  WW^{*}$, where $W^{*}W  =  I$ and $W$ is localised in the second.

(We could even choose $W$ in the same algebra $\mathfrak A(\mathcal O_1)$
if the latter were a so
called type III factor, which is most often the case by general theorems
 \cite{dopl5}).}, which   can just be assumed as an additional axiom besides duality.
Most of the analysis requires nothing more.

The collection  $\mathfrak A$ of quasilocal observables will be the operator 
norm 
closure of the union of all the $\mathfrak A(\mathcal O)$, that is, due to 
\eqref{dopleq1}
 and \eqref{dopleq2}, their 
norm closed inductive limit. Thus $\mathfrak A$ is a norm closed * subalgebra 
of 
$B(\mathcal H)$ (i.e. a {\itshape C* Algebra of operators} on $\mathcal H$) 
which is 
irreducible. The 
physical states of the theory are described by normalised positive linear 
functionals (in short: {\itshape  states}) of $\mathfrak A$, i.e. are 
identified with the 
corresponding expectation functionals.

A general comment on locality is in order: it is often claimed that the 
Einstein, Podolski and Rosen ``paradox'' shows that Quantum Mechanics is ``non 
local''. What does this 
statement mean? Folklore says that at least it does not mean that we can use EPR 
to transmit 
information. We would like to stress that certainly it is  {\itshape not } in 
contradiction with the notion of locality just recalled here. EPR shows 
that there will be states with {\itshape long range correlations}; but such states 
can 
be shown to exist in any theory which fulfils locality. In 
particular, the local algebras of free field theory provide 
mathematically precise sharp examples of this scenario.

Contrasts may well arise, however, between the EPR picture and a truly local 
picture of the measurement process \cite{dopl48}.

Unit vectors in $\mathcal H_0$ induce pure states all belonging to the same 
superselection sector, identified with the vacuum superselection sector; 
among these  pure states a 
reference vector state $\omega_0$, induced by 
the unit vector $\Omega_0$, will be called the {\itshape Vacuum State} (resp. the 
Vacuum State Vector).

In general, there will be a maze of other pure states (by the so called 
GNS construction, all appearing as vector states of other inequivalent 
irreducible representations of the algebra $\mathfrak A$).

 To implement the program outlined at the beginning of this section, we 
must define the ``suitably localised states'' in the theory. This will 
select, from all irreducible representations, those which describe 
superselection sectors (we must exclude the enormous family of 
mathematically possible but physically not significant representations, 
in the same way as in Quantum Mechanics we select from all the 
representations of the Heisenberg relations only those which are 
integrable into representations of the Weyl relations; but we must 
exclude also physically meaningful states which are not related to 
superselection sectors, such as the pure states describing the 
homogeneous equilibrium at finite constant densities and absolute zero 
temperature). In other 
words, we ought to consider representations describing {\itshape elementary 
perturbations of the vacuum}.

For the sake of simplicity of the exposition, we adopt here the 
restrictive notion of {\itshape double cone localisation} adopted in  
\cite{dopl6,dopl13} ;
 it was recognised  later (\cite{dopl7}, see also \cite{dopl9})  that the 
analysis goes through for a wider 
class ({\itshape spacelike cone localisation}) which was shown in \cite{dopl7} 
to cover all 
superselection sectors in any massive theory (but QED is left out by 
both).

A state $\omega$ is strictly localised in a double cone \(\mathcal O\) if the 
expectation value in  $\omega$ of any local observable which can be 
measured in the spacelike complement of \(\mathcal O\) coincides with the expectation 
value in the vacuum. In intuitive terms, we will select representations 
which, among their vector states, have sufficiently many strictly 
localised states, with all possible double cone localisations.

More precisely, it turns out that this is achieved by the following:

{\itshape selection criterion}:  the representations $\pi$ of $\mathfrak A$
  describing 
elementary perturbations of the vacuum are those whose restriction to 
$\mathfrak A(\mathcal O')$, for each double cone $\mathcal O$, is unitarily 
equivalent to the restriction 
to $\mathfrak A(\mathcal O')$ of the vacuum representation. This 
means that they describe 
``superselection charges'' which can be localised exactly in any tiny region 
of spacetime (note that an electric charge cannot be localised in this 
sense, as a result of Gauss theorem \cite{dopl6}).

It is important to note that such representations need not to be 
irreducible; the unitary equivalence classes of the  irreducible  
representations fulfilling the criterion will be the superselection 
sectors of the theory; their collection is thus determined by the vacuum 
sector together with the algebraic structure of the collection of all 
local observables.

Next step: up to unitary equivalence, the representations  $\pi$ of 
$\mathfrak A$   
fulfilling the criterion can be more conveniently described by ``localised 
morphisms'' of $\mathfrak A$  into itself.

For,  if the unitary operator $U$ implements the equivalence of $\pi$ and 
the vacuum representation when both are restricted to 
\(\mathfrak A(\mathcal O')\) for a chosen 
double cone \(\mathcal O\), we can realize the representation $\pi$ in 
question on the 
same Hilbert space as the vacuum representation, carrying it back with 
\(U^{-1}\). The representation $\rho$ we obtain this way is now the identity 
map on  \(\mathfrak A(\mathcal O')\):
\begin{equation} \label{dopleq4}
\rho (A)   =   A  \myif A \in  \mathfrak A(\mathcal O')
\end{equation}
and the duality postulate implies that it must map $\mathfrak A(\mathcal O)$ 
into itself; if $\mathcal O$ 
is replaced by any larger double cone, \(\mathfrak A(\mathcal O')\) is replaced by a smaller 
algebra, hence the forgoing applies, showing that any larger local 
algebra is mapped into itself; hence $\rho$ is an {\itshape endomorphism} 
of \(\mathfrak A\).

Since the choice of $\mathcal O$ was arbitrary up to unitary equivalence, our 
localised morphisms are endomorphisms of $\mathfrak A$  which, up to unitary 
equivalence, can be localised in the sense of (\ref{dopleq4})            in 
any 
double cone.

Unitary equivalence, inclusion or reduction of representations are decided 
studying their {\itshape intertwining operators} 
$T :   T \pi (A)   =  \pi' (A) 
T, A \in \mathfrak A$.  
Duality implies that the  intertwining operators between two  
localised morphisms must be {\itshape local observables}, in particular they 
belong 
to $\mathfrak A$.  Hence localised morphisms {\itshape act} on their 
intertwiners.

Now the composition of maps of two localised morphisms produces a 
localised morphism, their product; one can easily prove that:
\begin{quote}
{\itshape morphisms localised in mutually spacelike double cones 
commute}\footnote{\label{fn:morph}If  $\rho_n$ are morphisms unitary equivalent to  $\rho$  and localised in
double cones $\mathcal O_n$  which run away to spacelike infinity, the
unitary intertwiners
$U_n$ in $(\rho_n, \rho)$  are easily seen to induce automorphisms which
converge to $\rho$.
Choosing $\mathcal O_n$ contained together with $\mathcal O$ in some
double cone spacelike to the support of $\sigma$, we have that, for all 
$n$, $\sigma(U_n) = U_n$, hence in the limit  $\rho_n$  and $\sigma$ commute.

But this argument shows also that the "charged" state obtained composing
the vacuum state with $\rho$ is the {\itshape limit of vector states in
the vacuum sector}, which are bilocalised in  $\mathcal O$ and $\mathcal
O_n$, induced by the images of the vacuum vector through the $U_n$; these
states describe the limit state in $\mathcal O$ plus some compensating
"charge" in  $\mathcal O_n$ (we got a  {\itshape charge transfer chain}).
This gives a precise form to the old argument of the  {\itshape  particles
behind the moon} by Haag and Kastler  \cite{dopl1}.

If we replace the unitaries $U_n$ by their inverses, we can capture in the
limit the compensating charges localised in  $\mathcal O$ itself; they
will lie in another (  {\itshape  conjugate}) sector of the same family if
statistics, defined below, is  {\itshape  finite}.}.
\end{quote}
Composing a localised morphism with the vacuum state produces a state 
which is strictly localised, a vector state in a superselection sector if 
our  morphism is irreducible. Now we can define the product of such 
states! For, if $\omega_j  =  \omega_0 \circ \rho_j$, $j = 1, 2, 
\dots, n$ are such states and the morphisms used to create them from the 
vacuum are localised in pairwise spatially separated double cones, we can 
define
\begin{equation}\label{dopleq5}
\omega_1 \times  \omega_2 \times \dots \times \omega_n    
\equiv  \omega_0\circ 
\rho_1 \rho_2 \dots \rho_n
\end{equation}
a {\itshape product state}, 
since it will be independent of the order of factors 
thanks to the local commutativity of the localised morphisms, and will 
agree with $ \omega_j$ when tested with a local observable localised in a 
double cone which is spacelike to the localisation regions of all our 
morphisms except the \(j^{\text{th}}\) one.

This defines a commutative composition law among the classes of our 
representations, which can be interpreted as the {\itshape composition of 
superselection charges}; but the composition of irreducible representations 
might well be reducible.

(In mathematical terms, our localised morphisms $ \rho, \sigma, \dots$, and 
their 
intertwiners  $R \in (\rho, \rho')$ form a {\itshape tensor category}, 
tensor products of objects being the composition of morphisms, and that of two 
arrows, say $R \in (\rho, \rho')$,  $S  \in (\sigma, \sigma')$, being given 
by:
\begin{equation}\label{dopleq6} 
R  \times  S   \equiv   R \rho (S)   \in (\rho \sigma, \rho' \sigma').)
\end{equation}
Now the structure we described so far allows us to define STATISTICS.

If the morphisms in \eqref{dopleq5} are all equivalent to a given $\rho$, and 
say $U_j$ in $\mathfrak A$  are the associated (local!) unitary 
intertwiners, then the 
product 
$\rho_1  
\rho_2 \dots \rho_n$ is equivalent to $\rho^{n}$  and 

\[
U_1  \times  U_2 \times \dots \times  U_n  \in   
(\rho^{n},   \rho_1  \rho_2 \dots \rho_n ).
\]

Now obviously our states  $ \omega_j$  are vector states in the representation 
$\rho$  induced by the state vectors 
\[
\Psi_j   =   U_j ^{*} \Omega_0
\]
and we can define a {\itshape product state vector} $\Psi_1 \times  \Psi_2 
\times \dots\times  \Psi_n$ 
which induces the state $\omega_1 \times  \omega_2 \times \dots \times 
\omega_n$  in the 
representation  $\rho^{n}$ by setting:
\[
 \Psi_1 \times  \Psi_2 \times \dots\times  \Psi_n     \equiv   
(U_1 \times  U_2\times \dots\times  U_n )^{*}  \Omega_0.
\]
If we change the order $(1, 2, \dots , n)$ by a permutation $p$, the product 
state will not change but the  product state vector changes to
\begin{align*}
& \Psi_{p^{-1} (1)}  \times  \Psi_{p^{-1}(2)}  \times \dots \times  \Psi_{p^{-1}(n)}  =   \\
&(U_{p^{-1}(1)}  \times  U_{p^{-1}(2)} \times \dots \times  U_{p^{-1}(n)} )^{*}  
(U_1  \times  U_2 \times \dots \times  
U_n )  \Psi_1  \times  \Psi_2  \times \dots \times  \Psi_n  \\
&\equiv    \epsilon^{(n)} _{\rho} (p)  \Psi_1  \times  \Psi_2  \times \dots \times  \Psi_n .
\end{align*}
At first sight, we can only say that the unitary operator relating the two 
state vectors, the $\epsilon^{(n)} _{\rho} (p)$ {\itshape 
defined} by the last relation, belongs to the commutant of 
$\rho^{n}$. But it can be proved that:
\begin{enumerate}
\item 
 the map \(p \mapsto \epsilon^{(n)} _{\rho} (p)\) is a 
{\itshape representation of the 
permutation group} which depends upon $\rho$ only (not on the choice of the 
$U_j$);

\item 
if $\rho$ is changed to another localised morphism $\rho'$ by a unitary 
equivalence, say $U$ in \((\rho , \rho'), \epsilon^{(n)}_{\rho}\)  is 
changed to 
\(
\epsilon^{(n)}_{\rho'}\) by a unitary equivalence, implemented by  \(U\times U\times 
\dots\times U\);
thus the hierarchy of unitary equivalence classes of the representations  
$\epsilon^{(n)} _{\rho}$, $n = 2, 3, \dots$ depends only upon the unitary 
equivalence class of 
$\rho$, {\itshape a superselection sector} if  $\rho$ was 
irreducible.
\end{enumerate}
This hierarchy is then the {\itshape statistics} of that superselection 
sector.

The main result on statistics says that (as a consequence solely of our 
assumptions, that is essentially as a consequence of locality alone) the statistics of a 
superselection sector is uniquely characterised by a ``statistics 
parameter'' associated to that sector, which takes values  $ \pm 1/d$, 
or 
0, where $d$ is a positive integer. The integer $d$ will be {\itshape the order 
of 
parastatistics}, and + or $-$ will be its {\itshape Bose or Fermi character} 
(no 
distinction for infinite order, when the parameter vanishes).

More explicitly, let $\mathcal K$ be a fixed Hilbert space of dimension $d$, 
and let 
$\theta_n^{(d)}$ denote the representation of the permutation group of $n$ 
objects which acts on the $n$th tensor power of $\mathcal K$ shifting the 
factors; our 
theorem says that, given a superselection sector, if its statistics 
parameter $\lambda$ is $+ 1/d$ then, for each $n$, $\epsilon^{(n)} _{\rho}$ 
is 
unitarily equivalent to the sum of infinitely many copies of 
$\theta_n^{d}$; if $\lambda$ is $- 1/d$, the same is true provided we further 
multiply with the sign of the permutation; the latter being irrelevant if 
\(d = \infty\), i.e. if $\lambda = 0$.

In  mathematical terms this notion is canonical: the $\epsilon^{(n)}_{\rho}$ 
arise in a standard way from a ``symmetry'' for our tensor category (that is 
a map assigning to pairs of morphisms $\rho$, $\sigma$ a unitary intertwiner
\[
\epsilon ( \rho, \sigma)   \in ( \rho \sigma, \sigma \rho)
\]
which, in a precise mathematical sense, expresses the rule of  
commuting factors in the $\times$ product on arrows) which naturally 
arises here, 
since locality propagates to the arrows: $T \times S = S \times T$ 
if the sources of 
$T$ and $S$ are mutually spacelike localised morphisms, and the same is true 
for the targets; this symmetry is {\itshape unique} with the property that it 
reduces to the identity operator if $\rho$ and $\sigma$ are spacelike 
separated.

Can infinite statistics actually occur? The answer is yes in low 
dimension  \cite{dopl10,dopl11,dopl12} , where anyway the theory above does not apply: in $1 + 1$ 
dimension our category would not be necessarily symmetric, but only a 
{\itshape braided} category in general (a similar phenomenon occurs in $2 + 1$ 
dimensions in the case of the weaker spacelike cone localisation, see 
below); in $3 + 1$ dimensions, however, it can be proved that, in theories 
with {\itshape purely massive particles}, statistics is automatically finite; 
furthermore a slight generalisation of the above scheme (allowing 
localisation in spacelike cones - appropriate neighbourhoods of a string 
joining a point to spacelike infinity, suitable to describe topological 
charges) covers {\itshape all} positive energy representations and the whole 
theory can be extended to that case ( \cite{dopl7,dopl8}; see also 
\cite{dopl9,dopl13}).

Thus, in a widely general sense, to each superselection sector is associated 
(an integer, the order of parastatistics, and) a sign, \(+1\) for paraBose 
and \(-1\) for paraFermi. 

In relativistic theories, to each sector another sign is intrinsically 
attached, $ + 1$ for sectors with integer and $- 1$ for those with half 
integer spin values.

The Spin Statistics Theorem based solely on First Principles states 
that, for sectors with an isolated point in mass spectrum with finite 
particle multiplicity, {\itshape those signs must agree}.

This theorem, first proved for the class of sectors described here  \cite{dopl13} , 
was then extended to sectors localisable only in spacelike cones  \cite{dopl14} . 
More recent variants replaced the assumptions of covariance and finite 
mass degeneracy by that of ``modular covariance''  \cite{dopl15} . It has been 
generalised even to QFT on some appropriate kinds of curved spacetimes  \cite{dopl16} .

Note that the assumption  of finite multiplicity of one particle 
subrepresentations (either explicit, or, to some extent, implicit in 
the assumption of modular covariance: which in fact is based on the 
Bisognano Wichmann property, in turned proved, originally, for Wightman 
theories with finite tensor character) is an essential condition. For, as 
shown first by I. Todorov in the sixties, it is easy to construct free 
field models with the {\itshape wrong} 
connection between spin and statistics, yet 
fulfilling all the other axioms (Quantum Mechanics, Relativistic 
Covariance, Spectrum Condition, Locality), if we let in an additional 
{\itshape infinite dimensional} unitary representation of \(SL(2,C)\) acting 
on the 
internal degrees of freedom of one particle states. 
      
In a world with only one (or, for sectors which are only localisable in 
spacelike cones, with only two) space dimensions, as mentioned above, the 
statistics might be described by a braiding, not necessarily by a 
symmetry  \cite{dopl17,dopl18} ; the sign of the statistics parameter is replaced by a phase, as 
is the sign associated to univalence; again, in this general setting, 
these phases can be shown to agree  (\cite{dopl19,dopl20},  and References 
therein).

The connection between spin and statistics might fail also in a
nonrelativistic theory; necessary and sufficient conditions for its  
validity have been extensively studied \cite{dopl21}.

Dealing with a theory on the ordinary Minkowski space, but not necessarily 
assuming Covariance and Spectrum Condition, it is natural to restrict 
attention to {\itshape localised morphisms with finite statistics}, thus restricting 
the definition of  superselection sectors. They will be described by a 
tensor category of localised morphisms, where each object can be 
decomposed into a finite direct sum of irreducibles, and which possesses 
an additional important piece of structure.

Given two states as in \eqref{dopleq5}, the product state $\omega_1\times 
\omega_2$ 
need 
not to be a pure state even if the factors are pure, i.e. vector states 
in superselection sectors; it will however be at most a finite convex 
combination of pure states, vector states in some superselection sectors. 
Can one of these belong to the vacuum sector (so that there is a channel 
where the two factors can annihilate one another)? We would expect that 
this property characterises precisely sectors which are related by 
{\itshape conjugation of particle -antiparticle ``charge'' quantum numbers.}

Again, the general principle of locality allows us to {\itshape prove} 
that any 
superselection sector in our class has a conjugate in this sense (which
can be captured as mentioned in footnote \ref{fn:morph} with a careful use of the
charge transfer chains); more 
precisely, in mathematical terms, this allows us to prove that our 
category of localised morphisms with finite statistics is a ``rigid'' 
symmetric tensor C* category, where rigidity tells that to any object 
(localised morphism) we can assign another one such that the ``tensor 
product'' contains the tensor identity (the identity morphism)  as a 
component, with some minimality conditions which make its class unique  \cite{dopl6,dopl13} . 
To be slightly pedantic, the existence of conjugates identifies with 
rigidity if we first perform 
a trivial change, from the symmetry given by locality to another symmetry, 
changing the sign of its value on pairs of irreducible (para)Fermi 
morphisms.

Note that the identity morphism describes the vacuum sector, hence the 
tensor identity  is irreducible, that is its selfintertwiners reduce to 
the complex numbers. 

Now there is a wide, well known class of mathematical examples of rigid 
symmetric tensor C* categories with irreducible tensor identity: the 
unitary continuous finite dimensional representations of compact groups.

Any such categories has an additional feature, with respect to our 
superselection category: its objects are finite dimensional vector spaces 
(the representation spaces), and the arrows are subspaces of linear 
operators between the corresponding representation spaces (the ordinary 
intertwiners between the representations). The tensor operations here 
are ordinary tensor products of representations and of intertwiners, the 
flip of tensor products give the symmetry, and the complex conjugate the 
conjugation expressing rigidity.
 
Any rigid symmetric tensor C* categories with irreducible tensor identity, 
{\itshape provided it can be faithfully represented in the category of finite 
dimensional vector spaces}, 
is the dual of a unique 
compact group (by classical theorems of Tannaka and Krein). What in our 
more general case? Tannaka and Krein do not help, since our category is 
{\itshape not} represented in the required way.

This question called for a new duality theory for compact groups, where 
it was shown that each rigid symmetric tensor C* categories with 
irreducible tensor identity is the dual of a unique compact group, at the 
same time {\itshape proving} the existence of the desired faithful 
representation 
(as a symmetric tensor category) in the category of finite dimensional 
vector spaces  \cite{dopl22,dopl23}.

Here the main tool was the theory of highly noncommutative C* Algebras; in 
particular, it required the construction of a ``crossed product'' of a C* 
Algebra with centre reduced to the multiples of the identity (the algebra 
of quasilocal observables in our case) by a rigid symmetric full 
subcategory of endomorphisms (the superselection structure in our case); 
the automorphisms of this larger algebra which leave the original one 
pointwise invariant provide the desired (automatically compact!) group G. 
The cross product exists and is unique with some requirements, including 
the fact that each object $\rho$ of the category (the localised morphisms 
with finite statistics in our case) become inner in the larger algebra, 
in the sense that there are sufficiently many operators $ \psi$ there 
such that
\[
 \psi    A     =    \rho (A)     \psi,\quad         A  \in \mathfrak A
\]
If \(\rho_1,  \rho_2\) are irreducible and spatially separated, the 
corresponding 
\(\psi\)'s anticommute with one another if both the chosen sectors are 
paraFermi, they commute otherwise. Among other conditions, this is crucial to 
make the solution unique.

When $\rho$ runs through all our morphisms which are localised in a fixed 
double cone  $\mathcal O$ , these operators generate the local 
algebra of 
field 
operators in $\mathcal O$, and altogether, varying $\mathcal O$, the 
quasilocal field algebra \(\mathfrak F\); 
the set of fixed 
points in \(\mathfrak F\)  under $G$ is precisely $\mathfrak A$; 
the vacuum representation of $\mathfrak A$  induces 
the irreducible  vacuum representation of $\mathfrak F$, 
which restricted to $\mathfrak A$ gives a 
reducible representation, the direct sum of all superselection sectors 
each with multiplicity given by the order of parastatistics.  

In short we have constructed ordinary Bose/Fermi field operators, and 
the global gauge group acting upon them, whose irreducible representations 
label the superselection sectors, with dimensionality coinciding with the 
order of parastatistics. All compact groups must arise this way  
\cite{dopl24} .

Algebraic Quantum Field Theory provides also a weak form of the Noether 
Theorem, local current algebras  \cite{dopl25,dopl26,dopl27} , a 
generalised 
Goldstone Theorem  \cite{dopl28}, and 
allows us to discuss in mathematically precise terms the scaling limit and 
the phenomenon of confinement of superselection charges 
\cite{dopl29,dopl30,dopl31,dopl32}.

It is worth noting that the need for an abstract duality theory for 
compact groups, a problem which arose in 
Algebraic Quantum Field Theory at the end of the 
60's and was solved at the end of the 80's, emerged meanwhile in similar 
terms (for Algebraic Groups) in Mathematics, in the context of 
Grothendieck Theory of Motives; an independent solution, just slightly later 
 and with slightly different assumptions, was given by Deligne  
\cite{dopl33}. 
In recent years, Mueger gave an alternative proof 
of the Abstract Duality
 Theorem for Compact groups, following the line of the Deligne approach 
 \cite{dopl34}.

But what about the limitations imposed on the proofs of the connection 
between Spin and Statistics by the condition that only particles with 
positive mass appear in the Theory? And what about locality itself?

\section{The paradise lost: nonlocalisability of states and 
nonlocality of observables}

We pointed out in passing that electrically charged states will not be 
captured by the selection criterion described above (not even by its more 
general form in terms of spacelike cones). While the theory is believed 
(and indirectly checked, down to the scale of \(10^{-17}\;\text{cm}\) )
to be local, 
those states will not be localised, due to the slow decay of Coulomb 
fields \cite{dopl35,dopl36}. The relevant family of representations 
describing 
superselection sectors will have only asymptotic localisation properties.

It might still be, however, described by a tensor category of morphisms of 
our algebra of quasilocal observables; this category can at most be 
expected to be asymptotically Abelian in an appropriate sense; but this 
might well be enough to derive again a symmetry \cite{dopl37}.

We are still far from a Spin Statistics Theorem as described in the 
previous section, which applies to QED; however, in very general terms, 
it is reasonable to expect that it still holds when not only the charge 
carrying operations, but even the observables, are only asymptotically 
local, with sufficiently fast decays. For, in that case, scattering 
theory is still applicable, and if relativistic covariance holds, and if 
its validity propagates to the scattering states, the latter would be 
described by ordinary free fields, which must obey the connection between 
Spin and Statistics.

These comments however are far from being conclusive. Already in QED, the 
scattering theory becomes quite subtle \cite{dopl38} . But why should we 
worry about 
the possible breakdown of the very locality of observables? We are bound 
to face such a scenario if gravitational forces are taken into account.

At large scales spacetime is a pseudo Riemannian manifold locally modelled 
on Minkowski space. But the concurrence of the principles of Quantum 
Mechanics and of Classical General Relativity points at difficulties at 
the small scales, which make that picture untenable. For, if we try to 
locate an event in say a spherically symmetric way around the origin in 
space with accuracy $a$, according to Heisenberg principle an uncontrollable 
energy $E$ of order $1/a$ has to be transferred, which will generate a 
gravitational field with Schwarzschild radius $R \simeq E$ ($ \hbar = c =  
G  = 1$). Hence $a \gtrsim R \simeq  1/a$  and $a
\gtrsim 1$, i.e. in CGS units

\begin{equation}
\label{dopleq11}
a \gtrsim \lambda_P \simeq 1.6 \cdot 10^{-33} cm.
\end{equation}

However, if we measure one of the space coordinates of our event with 
great precision $a$, but allow large uncertainties $L$ in the knowledge 
of the other coordinates, the energy $1/a$ may spread over a thin disk of 
radius $L$ and thus generate a gravitational potential that would vanish 
everywhere as $L \rightarrow \infty$.

One has therefore to expect {\itshape Space Time Uncertainty Relations} 
emerging from
first principles, already at a semiclassical level. Carrying through such
an analysis \cite{dopl39,dopl40} one finds indeed that at least the 
following minimal restrictions must hold

\begin{equation}
\label{dopleq12}
\Delta q_0 \cdot \sum \limits_{j = 1}^3 \Delta q_j \gtrsim 1 ; \sum   
\limits_{1 \leq j < k \leq 3 } \Delta q_j  \Delta q_k \gtrsim 1 .
\end{equation}

Thus points become fuzzy and {\itshape locality looses any precise meaning}.
We believe it should be replaced at the Planck scale by an equally sharp and 
compelling principle, which reduces to locality at larger distances.

The Space Time Uncertainty Relations strongly suggest that spacetime has a
{\itshape Quantum Structure} at small scales, expressed, in generic units, by
\begin{equation}
\label{dopleq13}
       [q_\mu  ,q_\nu  ]   =   i \lambda_P^2   Q_{\mu \nu}
\end{equation}
where $Q$ has to be chosen not as a random toy mathematical model, but in
such a way that (\ref{dopleq12}) follows from (\ref{dopleq13}).

To achieve this in the simplest way, it suffices to select the model where 
the \(Q_{\mu \nu}\) are central, and impose the ``Quantum Conditions'' on the 
two invariants
 
\begin{equation}
\label{dopleq14}
Q_{\mu \nu} Q^{\mu \nu}
\end{equation}
\begin{align}
\left[q_0  ,\dots,q_3 \right]  &\equiv  \det \left(
\begin{array}{ccc}
q_0 & \cdots  & q_3 \nonumber\\
\vdots  & \ddots  & \vdots  \\
q_0 & \cdots  & q_3
\end{array}
\right)\\&\equiv  \varepsilon^{\mu \nu \lambda \rho} q_\mu q_\nu q_\lambda 
q_\rho =\nonumber\\&= - (1/2) Q_{\mu \nu}  (*Q)^{\mu \nu}\label{dopleq15} 
\end{align}
whereby the first one must be zero and the square of the second is $I$ (in 
Planck units; we must take the square since it is a pseudoscalar and not a 
scalar).

One obtains in this way  \cite{dopl39,dopl40}  a model of Quantum Spacetime 
which implements 
exactly our Space Time Uncertainty Relations and is fully Poincare' 
invariant. In any Lorentz frame, however, the {\itshape Euclidean} distance between 
two independent events can be shown to have a lower bound of order one in 
Planck units. Two distinct points can never merge to a point.  However, of 
course, the state where the minimum is achieved will depend upon the 
reference frame where the requirement is formulated. (The structure of 
length, area and volume operators on QST has been studied in full detail  
\cite{dopl47}).

Thus the existence of a minimal length is not at all in contradiction 
with the Lorentz covariance of the model; note that models where the 
commutators of the coordinates are just numbers $\theta$ , which appear 
so often in the literature, arise as irreducible representations of our 
model; such models, taken for a fixed choice of $\theta$ rather than for 
its full Lorentz orbit, necessarily break Lorentz covariance. To restore 
it as a twisted symmetry is essentially equivalent to going back to the 
model where the commutators are operators. This point has been recently 
clarified in great depth  \cite{dopl44}.

On the other side, a theory with a fixed, numerical
commutator (a $\theta$ in the sky) can hardly be realistic.

The geometry of Quantum Spacetime and the free field theories on it are 
{\itshape fully Poincare' covariant}. The various formulation of interaction 
between 
fields, all equivalent on ordinary Minkowski space, provide inequivalent 
approaches on QST; but all of them, sooner or later, meet problems with 
{\itshape Lorentz covariance}, apparently due to the nontrivial action of 
the Lorentz group on the {\itshape centre} of the algebra of Quantum Spacetime. 
On this point in our opinion a deeper understanding is needed.

One can however introduce interactions in different ways, all preserving 
spacetime translation and space rotation covariance; among these it is 
just worth mentioning here one of them, where one takes into account, in 
the very definition of Wick products, the fact that in our Quantum 
Spacetime two distinct points can never merge to a point. But it turns out 
that there is a canonical {\itshape quantum diagonal map} which 
associates to 
functions of n independent points a function of a single point, evaluating 
conditional expectation which on functions of the differences takes a 
numerical value, associated with the minimum of the Euclidean distance (in 
a given Lorentz frame!).

The ``Quantum Wick Product''  obtained by this procedure leads to a 
perturbative Gell'Mann Low formula free of ultraviolet divergences at each 
term of the perturbation expansion  \cite{dopl43} .

The common feature of all approaches is that, due to the quantum nature 
of spacetime at the Planck scale, locality is broken (even at the level 
of free fields, for explicit estimates see  \cite{dopl39}); 
in perturbation theory, its breakdown produces a non local kernel, which 
spreads the interaction 
vertices  \cite{dopl39,dopl41,dopl42}; 
this forces on us the appropriate modifications of Feynman rules
\cite{dopl40}. 

But  nonlocal effects 
should be visible only at Planck scales, and vanish fast for larger 
separations. If Lorentz invariance can be maintained by interactions, a 
point quite open at present, then we ought to expect the Spin and 
Statistics to remain true, as mentioned earlier in this section.

That argument might, however, raise the objection that, in a theory which accounts 
for gravitational interactions as well, there might be no reasonable scattering 
theory at all, due to the well known paradox of loss of information, if black 
holes are created in a scattering process, destroying the unitarity of the S 
matrix.

Of course, this is an open problem; but one might well take the attitude that a 
final answer to it will come only from a complete theory, while at the moment we 
are rather relying on semiclassical arguments. Which might be quite a reasonable 
guide in order to 
get indications of local behaviours; but scattering theory involves the limit to 
infinite past/future times; and it might well be that interchanging these limits 
with those in which the semiclassical approximations are valid, or with the 
infinite volume limit in which the thermal behaviour of the vacuum for a 
uniformly accelerated observer becomes an exact mathematical statement, is 
dangerous, if not misleading. And whatever theory will account for Quantum 
Gravity, it should also describe the world of Local Quantum Field Theory as an 
appropriate approximation.
 
One might expect that a complete theory ought to be covariant under 
general coordinate transformations as well. This principle, however, is 
grounded on the conceptual experiment of the falling lift, which, in the 
classical theory, can be thought of as occupying an infinitesimal 
neighbourhood of a point.  In a quantum theory the size of a "laboratory" 
must be large compared with the Planck length, and this might pose 
limitations on general covariance.

On the other side elementary particle theory deals with collisions which 
take place in narrow space regions, studied irrespectively of the 
surrounding large scale mass distributions, which we might well think of 
as described by the vacuum, and worry only about the short scale effects 
of gravitational forces. 

We are thus lead to consider Quantum Minkowski Space as a more realistic 
geometric background for Elementary Particle Physics. But the energy 
distribution in a generic quantum 
state will affect the Spacetime Uncertainty Relations, suggesting that the 
commutator between the coordinates ought to depend in turn on the metric 
field. This scenario could be related to the large scale thermal equilibrium 
of the cosmic microwave background, and to the non vanishing of the 
Cosmological Constant \cite{dopl45,dopl46}.

This might well be the clue to restore Lorentz covariance in the  
interactions between fields on Quantum Spacetime.

\end{document}